# A Method towards the Systematic Architecting of Functionally Safe Automated Driving- Leveraging Diagnostic Specifications for FSC design

**Authors:** Naveen Mohan, Martin Törngren, and Sagar Behere
Corresponding Author email: naveenm@kth.se

## Abstract

With the advent of ISO 26262 there is an increased emphasis on top-down design in the automotive industry. While the standard delivers a best practice framework and a reference safety lifecycle, it lacks detailed requirements for its various constituent phases. The lack of guidance becomes especially evident for the reuse of legacy components and subsystems, the most common scenario in the cost-sensitive automotive domain, leaving vehicle architects and safety engineers to rely on experience without methodological support for their decisions. This poses particular challenges in the industry which is currently undergoing many significant changes due to new features like connectivity, servitization, electrification and automation. In this paper we focus on automated driving where multiple subsystems, both new and legacy, need to coordinate to realize a safety-critical function.

This paper introduces a method to *support* consistent design of a work product required by ISO 26262, the Functional Safety Concept (FSC). The method arises from and addresses a need within the industry for architectural analysis, rationale management and reuse of legacy subsystems. The method makes use of an existing work product, the diagnostic specifications of a subsystem, to assist in performing a systematic assessment of the influence a human driver, in the design of the subsystem. The output of the method is a report with an abstraction level suitable for a vehicle architect, used as a basis for decisions related to the FSC such as generating a Preliminary Architecture (PA) and building up argumentation for verification of the FSC.

The proposed method is tested in a safety-critical braking subsystem at one of the largest heavy vehicle manufacturers in Sweden, Scania C.V. AB. The results demonstrate the benefits of the method including (i) reuse of pre-existing work products, (ii) gathering requirements for automated driving functions while designing the PA and FSC, (iii) the parallelization of work across the organization on the basis of expertise, and (iv) the applicability of the method across all types of subsystems.

## 1. Introduction

A modern vehicle can have more than 100 Electronic Control Units (ECUs) and implement about 300 different functions, which provide value to human drivers, across these ECUs. It is in this complex landscape that automated driving features are to be integrated. The level of control exerted by the vehicle can take many different forms based on use cases, operational scenarios and capabilities of the vehicle. Through all the varying forms of automation, an OEM needs to be able to assure that the developed function can perform at least as well as the driver in the modes in which the function takes over control while providing adequate functional safety. While automotive systems design has traditionally been an iterative bottom-up approach, newer functional safety standards such as ISO 26262 [1], the de-facto standard for the automotive industry, prescribe a top down approach.

In the development process prescribed by ISO 26262, functional elements from the Preliminary Architecture (PA) of the Functional Safety Concept (FSC) are broken down and detailed successively with the progression of the design phase. However, if this is done without considering underlying elements of the platform, it could lead to solutions with extensive development expenses. While cost is not the most important concern for safety critical system design, within the cost sensitive automotive industry there are significant gains to be made in reuse of components and technologies [2]. ECUs and systems are usually a product of several years of stringent development. It is thus probable that safety critical subsystems will be reused entirely and with the explicit intention of keeping changes as minimal as possible for the new functional needs. Hence, most architects will keep platform considerations in mind while dealing with design and lean towards the reuse of proven components from their platform. The challenge of architecting the PA thus transforms from "given these requirements, what do I need to design?" to "given what is available from the platform, what is the minimum that needs to be designed additionally to satisfy these requirements?"

To answer the question architects have traditionally relied on experience. However, for this tacit knowledge to be used in the design of safety critical systems developed under the stringency of ISO 26262, the platform considerations need to be extracted and addressed explicitly. Based on their experiences from design of Functional Safety Concepts for L3 and L4 automated functions using the design method in [20], the authors of this paper advocate the incorporation of platform information as early as the FSC sub-phase. This paper aims to expand this idea and address the challenge of which information is to be extracted and how, using the following research question

**RQ: How can detailed domain specific information about legacy subsystems be best extracted from the platform for the purposes of design of the FSC?**

The paper addresses this question in two steps, first by defining heuristics to analyse the influence of human vehicle drivers (from here on abbreviated as "driver") on the subsystem, in the form of a structured method. The second step is to collect the information generated by the domain expert's application of the method in a subsystem report. The subsystem report serves as a basis for architects to make their decisions. The paper further applies this method as part of an industrial case study using one of the most critical subsystems in the vehicle, the service brakes, and demonstrates the effectiveness of the method in providing relevant critical information to the system architects.

The method thus aims to provide a well-defined baseline of information from the platform, about a class of functions related to a heuristic, generated by multiple domain experts in parallel for the vehicle architects.

The rest of this of this paper is organized as follows: Section 2 will shortly explain the background and terms used through this paper. Section 3 explains the conceptual approach for the method in Section 4. The case study is described in Section 5 followed by a discussion in Section 6. Section 7 gives an overview of the related work and Section 8 gives the conclusions.

## 2. Background

### 2.1. Terms

**ADI, Platform and Legacy**

A *platform* is defined here as the set of *components* that provide Electrical/Electronic (E/E) functionality for a vehicle including (but not limited to) hardware and associated low-level control. Components include physical parts such as communication networks, power networks, ECUs, sensors and actuators etc.; and functional components with which we refer to individual software components and logical aggregation of these components into functions. Different product lines share components from the platform and products differ only in their composition of components used.

The term *legacy*, used as an adjective, is used as a temporal qualifier. All components that are available at the time of initiation of a new project are legacy components. Thus, the platform available at the time of initiation of a new project is a legacy platform, and the components therein are legacy components.

To distinguish the intelligence needed for the task of driving automation from the components in the platform, we use the term *Autonomous Driving Intelligence (ADI)*. The ADI is defined as those parts of the Automated Driving System defined in the J3016 [3], which are not part of the legacy platform. The notion of the ADI is useful for reasoning about the increased functionality of the vehicle, by dealing with the ADI as an addition to the platform.

**Subsystem and Architecture**

A *subsystem* is a set of components that comprises an ECU, its software, the sensors that provide it input, the actuators that it controls and the interconnections between them.

Architecture is defined by ISO 42010 as "fundamental concepts or properties of a system in its environment embodied in its elements, relationships, and in the principles of its design and evolution" [4]. ISO 26262 however takes a more tangible definition of the architecture as the *"representation of the structure of the item (1.69) or functions or systems (1.129) or elements (1.32) that allows identification of building blocks, their boundaries and interfaces, and includes the allocation (1.1) of functions to hardware and software elements 1.4"*. This paper uses the definition from ISO 26262 for the term *architecture*.

The *Preliminary Architecture* is the architecture used within the Functional Safety Concept subphase and is used for allocation of Functional Safety Requirements (FSRs).

**Organizational Roles Referred To in this Paper**

Throughout this paper, the term *architects* refers to architects at the vehicular level, unless specified otherwise. By *safety engineers* we refer to role tasked with assuring functional safety of the product. By *domain experts* or simply *experts,* we refer to a role providing expertise in specific areas and technologies.

Architects and domain experts have different responsibilities. The architect needs to know how a particular component will be used at a vehicular level both currently and in the future while the domain expert typically knows in depth the capabilities and limitations of a particular subsystem.

The safety engineers are responsible for the FSC work product while the architects provide a PA of sufficient granularity as an input and consult the domain experts for domain specific expertise.

**Levels of Automation**

The terms L0, 1..5 are defined in the SAE standard J3016 [3] which classifies automation in driving into six broad levels, from Level 0 (no automation) to L5 (full automation under all driving modes).

**Method**

To distinguish the context of the usage of the term from other conflicting definition, we use the term *method* as defined by Estefan in [5] in that a method *"consists of techniques for performing a task, in other words, it defines the "HOW" of each task.* Methods can be prescriptive or descriptive.

## 2.2. Diagnostic Specifications

Diagnostic specifications have long been a standard work product in the automotive industry for both purchased and inhouse developed subsystems. An overview of diagnostics in automotive domain and how they are used can be found in [6]. Though the original intention behind diagnostic specifications was to record Diagnostic Trouble Codes (DTCs) that were used to assist in workshop repairs, diagnostic specifications have since evolved to include details of all monitors used in a particular subsystem and are used extensively for troubleshooting even during development. A diagnostic specification typically contains a list of diagnostic monitors used in the software; trigger conditions, descriptions of how each monitor works; healing conditions and system reactions, should a monitor be triggered. ECUs could also save additional information, known as snapshots or freeze frames, along with the DTCs to assist in the diagnosis of the failure. Diagnostic monitors are used to trigger failure related notifications to the driver. A common way of alerting the driver about failure and need of service is by the use of warning lamps. In this paper, we use a basic classification of the lamps and distinguishing only between red and yellow lamps. A red warning lamp usually indicates a critical failure that renders the vehicle unsafe and requires an immediate stop whereas a yellow warning lamp is generally associated with a failure that is severe but does not require immediate stop in the vehicle. It is an established state of practice for OEMs to maintain detailed diagnostic specifications and related repair information per component even if the component in question has been purchased from a supplier.

Since the failure detection capabilities at a subsystem level rely primarily on these diagnostic monitors, they provide an ideal point of reference for safety related work as well. An additional advantage is that failures in the field are usually tracked using DTCs and thus most OEMs have developed extensive infrastructure to gather information about failures using DTCs. Though not in the scope of this paper, this can, in the future, give statistical insights into the qualification of components, for use in safety-critical vehicular functions.

## 2.3. Influence of the driver on the automotive systems design

Vehicles have traditionally been developed solely for use by human drivers. Thus, it is not surprising that the influence of a driver has permeated the design choices made during vehicular systems. The influence of a driver is apparent in subsystems which directly interact with the driver such as an infotainment subsystem. A platform also contains several interfaces to driver such as brake pedals and steering columns that could be deemed unnecessary in an automated driving context. But the influence of a driver is also seen in subtler ways in the design of other subsystems as elaborated in the following paragraphs.

The design of diagnostic monitors for functional safety can be influenced by the presence of the driver. ISO 26262 for example uses the notion of controllability which allows for reduction of the harm associated with a hazardous event based on the ability of the driver (or others at risk) to control the hazardous situation once it has occurred. A lower ASIL level could be assigned for a failure that could be noticed reasonably quickly by a driver assuming the hazard was controllable. Extending this notion, for certain cases, it could be reasonable during a development process to emphasize on *detectability* of fault rather than *avoidance* of failure, if the driver can be made aware of the failure very quickly.

Other sources of influence of driver on design could include trade-offs being made for the driver's comfort. For example, the low frequency hum of an electric motor might necessitate a reduction in efficiency or a complicated design change solely for the comfort of the driver.

## 2.4. ASIL tailoring

ISO 26262 mandates a strong linkage between the requirements related to safety to the architectural elements at every abstraction level. The decomposition of safety requirements, and in particular the so called ASIL decomposition depends on exploiting sufficiently independent architectural elements, each capable of fulfilling a given safety requirement, to reduce the ASIL level of the original requirement by demonstrating suitable redundancy with the partitioning of the system. ASIL decomposition can be applied to the functional, technical, hardware or software safety requirements i.e. at any abstraction level, given appropriate architectural elements and adequate information about them.

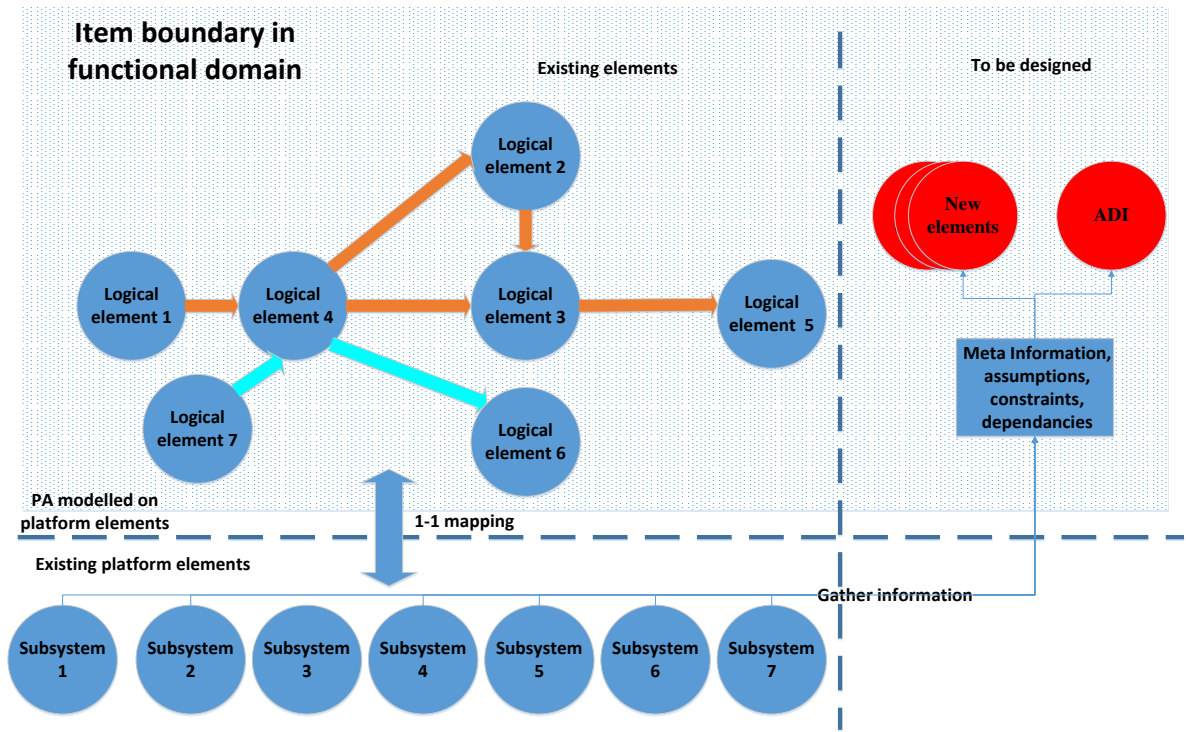

**Figure 1: Gathering of information from subsystems with the PA elements based on subsystems**

ASIL tailoring refers to the effective allocation, using decomposition if needed, of safety requirements to match the elements of the architecture at that level. Though ISO 26262 does not recommend any particular phase for ASIL tailoring, this paper argues for decomposition to be performed as early as possible and preferably within the FSC.

## 3. Conceptual Approach

Automation in driving can be architected in many different ways. An architect has to decide which constellation of components from the platform fit together and which need to be developed anew to fulfil the company's technology roadmap.

ISO 26262 does not contain specific guidelines for the level of abstraction of the elements of the PA. The guide word used in the description i.e. "functional" is not suitable as functionality can be abstracted at many different levels. From the authors' experiences in designing the FSC for a driving function with an L4 level of automation, it was found useful to (i) to use an abstraction of ECU systems as elements of the PA (ii) with the failure modes being defined as the unavailability of function. The advantages gained were

- by early incorporation of platform considerations, the FSC required less rework in the later phases and the subsequent architectural decomposition could iterate to the platform components faster.
- It allowed for the tracking of constraints between subsystems and to help judge implications of using them
- it affords an easy way to connect failures at the component level, which are usually well known and documented, to failures at the vehicular level.
- The knowledge of failures of the PA elements allows for analysis and inference of independence of the elements and paves the road for decomposition of FSRs.

Fig 1 serves to *illustrate* the approach using a one to one mapping between the platform components and elements of the PA. The thick arrows indicate functional interaction, and consequently constraints between the subsystems. For the purposes of defining the problem faced by architects during PA design, the entire platform is envisioned to be divided into three different divisions (i) the subsystem under consideration, (ii) the other subsystems in the platform and the (iii) ADI.

Using this model, the question posed in the introduction can be broken down into:

a) Given the selection of a particular subsystem i.e. (i), what are the restrictions placed on the selection of other subsystems i.e. (ii) and the ADI?
b) Given the answer to a), what new elements must be designed along with the ADI.

To answer a), the domain expert's information about the subsystem thus needs to be abstracted to the architect's level in a suitable way. A challenge in using in the traditional ways of organizing meetings to gather information as needed is scalability. A consequence of the diversity of the subsystems being part of the item is that multiple experts need to be consulted by the architects. This is unavoidable as the redundancy required for ASIL decomposition could come both from within the subsystem or without.

During their design of the FSC for L3 and L4 functions, the authors noted that there were common classes of information required from every subsystem involved. The lack of this information, resulted in discussions with a number of experts about similar issues, and was repetitive and time consuming. While some of this information will remain situation dependent, it was judged that there could be significant reductions in the time taken for FSC design if the common information could be assimilated at the beginning of the task.

With the varied nature of subsystems in the platform and functions, it is difficult to extract information for every scenario with one specific method, since the influence on the functions at vehicular level requires 1) the context in which the particular failure will occur, 2) information that is not typically part of the domain expert's responsibilities. The authors' postulate that by defining a class of functions and using heuristics organized for a particular class into a method, much of the information needed can be obtained from a domain expert, without involvement from the architect.

Due to the inherent connection between the diagnostic specifications and the failure modes of the subsystem, their ubiquity in automotive subsystems, and due to their connection to field failure data, the selection of the diagnostic specifications as a target for the heuristics was a natural choice.

The heuristics for the class of automated driving are centred towards the identification, and hypothetical removal of the influence of the driver from the vehicular system design. Using an existing work product viz. the diagnostic specification of a subsystem, the method leverages it as a basis to communicate a subsystem report to the vehicle architects at the required abstraction level for the FSC. The resulting subsystem report gives information about what the subsystem's capabilities and limitations. The order in which the heuristics are applied in the method were devised as a result of iterating with a toy example and tuned to generate the type of information required at the right level of abstraction.

To summarize, it is preferable to decompose safety requirements as early as possible in the safety lifecycle. The conditions to decompose safety requirements depend heavily on architectural information and the proof of "sufficient independence" of the elements used. In the case of platform based components, this information can be obtained from the domain experts. An architect can use the information about different subsystems to provide appropriate architectural granularity for the safety engineer to use. In the information that needed about the particular subsystems, there are certain common classes of information that if collected prior to starting the design process, shortens the FSC design time and helps avoid repetitive tasks. The method proposed in section 4 is designed to collect the information for a class of functions related to automation using heuristics to give a basic context to the domain expert. The report generated by the domain expert is utilized to make architectural decisions about the elements of the PA and assist the safety engineer.

## 4. Proposed Method

This method is to be followed by all subsystems that the architect identifies as necessary for functions related to driving automation. The method comprises 10 heuristics, constituting questions. The subdivisions marked with lettering a., b. etc. are individual steps in their own right while the bulleted lines are informational to illustrate the purpose of the steps. The method takes the diagnostic specification of a subsystem as input, and delivers a subsystem report as the output. The method is designed to be applied by the domain expert without interference from the architect. The responses contain, per monitor, the initial information from the diagnostic specification and the responses to each question.

1. Mark all diagnostic monitors that detect failure with parts of the subsystem that physically interact with the driver
- Rationale: These are likely redundant if a driver is not present i.e. automated functions of SAE level L4 or L5.
- E.g. Monitors related to the foot brake module will become redundant if the foot brake module is removed.
    a. Identify the preferred new interface needed for communication of the same information to the subsystem from the ADI.
    - Rationale: components with human interaction usually act as sources of input to subsystems. The alternative source of input i.e. the ADI will need an interface to communicate electronically.
    - E.g. The ADI needs to be able to request braking torque in an appropriate way. It will likely need a message on the network at a certain frequency and granularity in requested torque.
    b. Are there known issues in providing these interfaces electronically?
    - Rationale: Identify limitations in legacy.
    - E.g. Legacy expectations on frequency of input.
    - E.g. CAN load on the network.
    c. Does the electronic control improve the performance of the subsystem or reduce it?
    - E.g. faster or more accurate setpoints for braking are possible with the ADI.
2. Investigate the influence of assumptions of a driver in the design of diagnostic monitor. Identify
    a. If the current design of the monitor has been based on perceived feel of driving by the driver.

- Rationale: Identify diagnostic monitors that solely record symptoms a driver might encounter for the purposes of troubleshooting field reports.

  b. If time for transition to failsafe mode needs to be changed and note limits to the change.

  - Rationale: This information can be used as a measure of responsiveness for the system
  - Faster transition may be possible because driver is not present and will not be affected by the change.
  - Slower transitions could be possible since greater control can be exercised by the system compared to the driver.

  c. Make a short summary of how this failure may influence functions at vehicle level.

  - Rationale: The failure may lead to an indication to another subsystem if not the driver. This could potentially be tracked by the ADI.
  - E.g. A slave subsystem may indicate a failure to a master by signalling a lower quality of service.

3. Mark diagnostic monitors that indicate a need for failure handling at a vehicular level i.e. beyond the control of the subsystem in question

   - Rationale: These monitors identify potential gaps in FSC if this subsystem is included in the design. These gaps will need to be handled by the ADI or by the creation of a new safety mechanism.
   - E.g. Power supply related failures
   - E.g. Secondary failures or markers which indicate missing signals from other ECUs etc.

4. Identify changes that assist of detection, notification and reconfiguration using monitors that use warning lamps to alert the driver. Suggest new interfaces if necessary.

   - Rationale: If the subsystem signals the driver, this must be a significant failure and should be considered at a vehicular level if it cannot be handled with changes to the subsystem.

     a. Target monitors which set yellow warning lamps for intra subsystem reconfiguration. Is there a degraded mode that is possible for the subsystem to achieve with redundancy within the subsystem?

     - E.g. redundant brake circuits.

     b. Target monitors which set red warning lamps for inter subsystem reconfiguration; mark other existing subsystems that can fulfil some parts of the function.

     - E.g. Parking brakes could fulfil partial functionality of the service brakes
     - E.g. Speed estimation could be obtained from either the transmission or brake subsystems, even if precision might vary.

5. Investigate role of driver in failure detection.

   - Rationale: Reliance may have been placed on detection by driver

     a. For each monitor note the information that is needed to diagnose the failure.

     b. How are these failures currently detectable by a driver if they occurred?

     - E.g. sounds or jerks in the vehicle?

     c. Has the decision on monitor criticality been influenced by detectability by drivers?

     d. How can the failures be detected by the ADI instead? Note which type of data is needed and its format.

     - E.g. CAN signal, ECU memory?

6. Determine influence of driver in latent fault detection. Are there any tests for safety currently performed by a driver for the subsystem?

   - Rationale: Checks currently performed by driver for latent faults must still be performed.

     a. How can these tests be performed electronically? Can these be performed only at start of drive cycle or do they have to be performed regularly during drive.

     - E.g. tyre pressure checks could require a short drive before mission start.

7. Trace propagation of failure. Which other subsystems are affected if the failure detected by the diagnostic monitor occurs and how do they react if
   a. The failure is detected by the monitor
   b. The failure is not detected by the monitor

8. Check the effects of variability. Go through all the findings for determining a minimum configuration of the subsystem

   - Rationale: Not all variants of the subsystem may be ideal for automation
   - E.g. automated transmissions are easier than manual transmissions for control by the ADI.

     a. Is there a minimum configuration needed for this particular subsystem in the absence of the driver?

- Which configuration implies lower number of failures that affect other subsystems?
   b. What impositions are made onto other subsystems on the basis of such a selection
      - What other subsystems need to be present in the vehicle if this minimum configuration is chosen. i.e. map dependencies to other subsystems
9. Where possible, generate requirements that the ADI must fulfil if this subsystem is to be used based on the work so far.
10. Where possible, look up field data to derive frequency of each failure.

The summarized responses for these 10 steps, applied to the relevant subsystems, are then submitted to the system architect.

# 5. Industrial Case Study

The study was conducted as part of project ARCHER, between Scania C.V. AB and KTH Royal Institute of Technology, which deals with functionally safe architectures for fully automated driving. The service brake subsystem was chosen for the case study by virtue of it being an essential safety-critical component in every automated driving scenario, and since it was likely to be reused entirely. It was thus seen as ideal to showcase the method.

## 5.1. System Description

A modern truck is usually equipped with multiple ways to brake which exhibit different performances. European regulations require a truck to be able to brake smoothly to a stop in completely redundant ways. Figure 2 provides an overview of the different types of braking available in the platform. The platform from the OEM at the time of choosing offered the following possibilities:

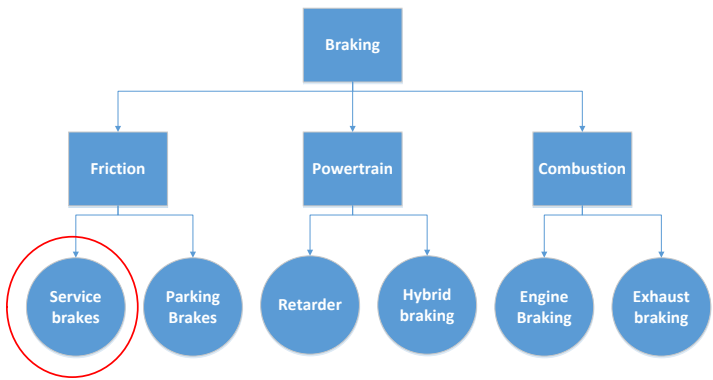

**Figure 2: Overview of braking technologies available in the platform**

i. **Service brakes:** These are the most common brakes that are mounted in a vehicle and use friction exerted on the wheels to brake the vehicle.
ii. **Parking brakes:** The parking brake also uses friction on the wheels to slow the vehicle down. Though the main purpose of the parking brakes is to keep the vehicle at standstill, in many cases it is used as an emergency fall back solution, should the service brakes fail. It can be controlled either mechanically or electronically depending on the variant selected.
iii. **Engine brakes:** This encompasses the various types of braking forces that can be exerted via the appropriate control of the combustion engine in a vehicle.
iv. **Exhaust brakes:** Braking is achieved by constricting the exhaust pipe using valves, to increase back pressure, thereby influencing the combustion and consequently slowing the vehicle.
v. **Retarder:** This method of braking is hydraulic in nature and works by pumping a fluid thorough the transmission chamber to exert a braking torque via the viscous drag induced by the fluid.
vi. **Hybrid braking:** In this case, braking is achieved by active control of the electric motor in a hybrid drivetrain to exert braking torque.

Each of these braking subsystems could have multiple variants in software or hardware to account for strength of braking, number of wheels and axles, market specific legal requirements, business needs etc. The use of (i) and (ii) is legally mandatory for vehicles in Europe and every vehicle is equipped with a variant of these. (iii) and (iv) are commonly in place in most vehicles while (v) and (vi) are optional features.

For the purposes of this study we focus on technology (i) i.e. service brakes, which are implemented at Scania using a subsystem called the Electronic Braking Subsystem (EBS). The EBS subsystem, in addition to controlling the friction brakes, has the responsibility to interpret the driver's braking request from the brake pedal. EBS acts as the master to all the other subsystems for braking related functions and allows them to actuate only under suitable conditions. EBS is also responsible for various functions related to vehicle dynamics and stability such as the ESP, ABS etc. At Scania, the EBS is a subsystem bought in its entirety from a Tier1 supplier. It actuates a complex pneumatic system that uses dual channel circuits to control multiple Pressure Control Modules (PCM) (one per wheel) to deliver the required amount of braking per wheel. It also contains other components common to pneumatic braking systems such as reservoirs, pressure regulating modules, different sensors to measure performance including temperature, load and speeds etc. The EBS software is capable of handling variants in hardware using parameterization performed at the point of manufacture or in the aftermarket. Fig 3 shows the internal composition of the EBS subsystem.

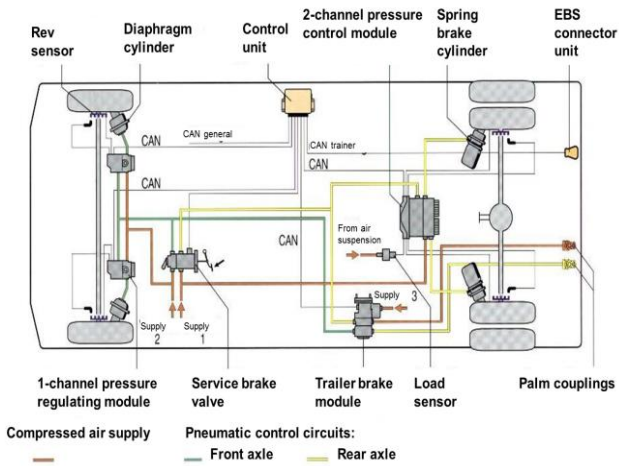

**Figure 3: Internal composition of EBS subsystem**

## 5.2. Case Study and Results

The case study involved the application of the methods discussed in Section 4, to the diagnostic specification of the EBS subsystem. Due to conflicting schedules with the EBS domain expert, this study was conducted with the help of the architects who were in charge of EBS and vehicle dynamics domain, as this was deemed to be sufficient for the purposes of evaluating the method.

The diagnostic specification was quite extensive and was analysed manually. The version of the diagnostic specification that was analysed was an excel export of the suppliers diagnostic data management system.

The export comprised 700+ unique rows that represented unique checks of the software on different elements of the subsystem and collectively, there were about 210 unique DTCs that could be set by the monitors. Thus monitors and DTCs were asymmetrically mapped in that some DTCs could be set by multiple monitors and other monitors did not have a DTC linked to them. This was expected as DTCs have been traditionally optimized to highlight components for replacement to a workshop technician, and not necessarily for diagnosing faults with precision. Each of the monitors was annotated to show a) which functions of the EBS subsystem such as ABS, lining wear control etc. would be affected, b) which of the two distinct warning lamps would be used to indicate failure to the driver and c) possible causes of the failure and repair actions.

Due to the exhaustive nature of the work, only the important findings from the case study related to nuances in applying the method and the impact on the architecture are discussed here for the purposes of demonstrating the usefulness of the method. The overarching emphasis of this study was on deciding which failures the ADI can detect, determine what it should know if it cannot detect that particular failure, and addressing of these cases with the use of new interfaces and software components. Thus that is the example that will be expanded upon primarily, in the following paragraphs.

Each of the monitors were analysed by the method to determine the role of the driver and parts of the subsystem such as the FBM which dealt with driver interaction were marked to be not necessary for a scenario with no driver. The EBS subsystem already contained an interface for electronically receiving braking requests and hence a new interface to the subsystem was not necessary. In other cases such as monitors detecting if a valid steering angle was available over CAN, alternate sources of the same information from the platform were identified and interfaces to the EBS were suggested. Some monitors were highlighted for possible improvement in the cases where the driver was part of the detection process. E.g. a continual increase in requested braking torque with the foot pedal was used to detect the omission failure mode for the EBS or one of its slave subsystems i.e. the driver did not get the retardation needed even if the subsystem thought it was being delivered. A monitor like this could be reused by the ADI mimicking a driver by making estimations on speed and position based purely on the environment, independent of the platform components, before requesting retardation from the EBS. This was preliminarily formulated as a requirement on the ADI. Variants in the subsystem and the dependencies induced by them to other subsystems were particularly tough to handle. A decision was made the authors, to start with the variant that was most feasible to automate and progressively add other variants into consideration, at later stages of the project.

During the study, in cases of uncertainty caused by a lack of detailed expertise, reliance was placed on the initial annotation of warning lamps. Monitors that would only set yellow warning lamps when triggered were deemed to be not safety critical in the current drive cycle, and monitors that set red warning lamps were assumed to be relevant to functional safety within the same drive cycle. This decision was made as the monitors that set yellow lamps were found to be serving as indicators for failures that had not yet occurred i.e. as indicators for failures that could occur in the future if the tolerance limits for the monitor were violated and not immediate failures. These "yellow" failures could be thus detected as part of startup tests or before automated driving modes would be activated in the vehicle. Essentially an assumption was made that the system had the capability to refuse to accept missions in automated driving modes until a certain number of tests had been performed including the startup related tests.

The unavailability of the function was judged to be more important to the ADI rather than which particular cause was responsible for the unavailability. Thus a decision was made that availability of functions should be information that the ADI should be made aware of. By temporarily removing monitors marking faults in other parts of the vehicle e.g. CAN time outs etc., number of monitors for consideration to about 500 and it was noted that this information needed to be transmitted to the ADI. Only about 330 were considered to be capable of affecting immediate harm in a short time frame using the warnings lamp type heuristic and removing monitors marked with yellow from consideration. Trailer related monitors were eliminated from consideration if they did not directly influence the capacity of the tractor. Foot Brake Module related monitors were also excluded as the ADI would control the vehicle using signals over CAN and not analog signals. Taking into consideration the symmetry in the vehicle i.e. repetition of monitors per wheel, the number of monitors to be considered were reduced to 60. These were examined to contain approximately 20 startup DTCs that could be detected within current methods and needed no modification. Out of the remaining 40 monitors that detected failures not possible to handle within the subsystem, each monitor was used to generate requirements on the ADI to fulfil.

To ensure backwards compatibility with existing manual variants, the possibility of adding a new Software Component (SWC) to act as a gateway between the ADI and the EBS subsystem was considered and found to be promising at this stage of analysis. This SWC could be allocated to the EBS ECU such that it could access the same information that was physically available to the other SWCs and minimize the changes to the other SWCs.

To summarize, the subsystem report generated as part of the method gave deep insights into the working of the EBS subsystem, highlighted the dependencies on the ADI and other subsystems that needed to be considered, if the subsystem was used. The subsystem report was judged to be promising in solving that were encountered in the FSC design by the architects and safety engineers.

## 6. Discussion

### 6.1. Perceived Merits

Claiming compliance with ISO 26262 implies maintaining a strong connection between requirements and architectural elements through all the phases of development. Functions related to automated driving require the use of a large number of subsystems and it becomes a significant task to keep track of the diverse design information. The impact of the varying types of information is very evident in the early phases of design where there is uncertainty in the both functional requirements and operational scenarios. Since these uncertainties depend on several qualitative aspects such as stakeholder negotiation, even the process of gathering the required information becomes hard to formalize without unrealistic assumptions.

Given that the elements of the PA need to map to the platform components as much as possible for the purposes of cost savings, incorporating the knowledge of platform components early in the process has the potential to save several iterations in design and the cost associated with following the rigorous process induced by ISO 26262. In this context, the existence of components with years of design, and service in the field can greatly help in reducing the uncertainties in design by constraining the design space. This paper exploits the use of the information available to speed up the design of the FSC sub-phase. In doing so, the important dependencies between subsystems and the ADI are traced, potentially shortening the time for development for the ADI and the new elements that need to be designed. This task is accomplished by the heuristic based method suggested in the section 4. An advantage of the heuristics based method and using the diagnostic specification is that it was perceived to be simple to understand by the organization, and lead to a straightforward application to every subsystem. This information greatly reduces the amount of communication between the architects and the experts as this is done only once per class of functions.

The method gives a separation of concerns using relatable heuristics to provide basic usage context, i.e. the removal of the driver, allowing for the domain expert to work independently of the architect and be agnostic of the specific usage context for the particular subsystem. Thus, the method addresses another common issue in the design of automotive systems in that a component may be used in various functions at a vehicular level and it is hard to predict the behaviour of the component at this level without in-depth information about context, operational scenarios etc. The architect is benefitted by the information and is able to judge which of the information generated is directly applicable to his design and has the agency to ask more detailed questions to the domain expert if needed, using the subsystem report as a basis for consistent discussion.

The case study was intensive in terms of man hours, however the authors estimate that this is due to the selection of one of the most challenging subsystems to showcase the method and that it should be easier for other subsystems that are simpler in design.

Though the focus of this paper is on safety and not on other extra functional properties such as availability, it lays the ground work for reasoning about the other extra functional properties to be made at the vehicular level in the future. The subsystem report also helps guide how a subsystem needs to be modified in the absence of a driver. Information such as dependencies between variants of subsystems etc. allows an architect to better judge the impact of modification of the particular subsystem and decide where a new subsystem will be needed instead. The subsystem report from the case study was found to be useful for the purposes of design for automation related functions by architects and safety engineers.

The method aims to connect properties of the underlying subsystems known to the domain experts, which might otherwise not have been considered, to vehicle level properties for use by architects. The use of diagnostic specifications is beneficial for functional safety design as this enables the use of information from the aftermarket about failures and consequently statistical analysis of failure rates of the subsystems. Such a link will be crucial during the verification phase of the FSC.

A claim of limited generalizability of this method can be made due to its broad applicability by virtue of depending solely on a commonly used work product (diagnostic specifications) and domain expertise. The method works on subsystems irrespective of whether they were developed in-house or at suppliers. Since the subsystem report output delivers useful information at the right abstraction level, and is able to be tailored to organizational needs, the method was judged to have shown progress in answering the research question posed.

## 6.2. Limitations and Scope

This paper makes several assumptions based on standard practice, in the development of automotive systems such as role definitions and the argument for platform based design etc.. While these assumptions need not be consistent across every organization to the same degree, care has been taken in this work to ensure that only realistically generalizable assumptions were made.

Diagnostic specifications are a standard work product across the OEMs and are used in one form or another to track issues in the aftermarket. By using only a small and most commonly used subset of the possibilities offered by the diagnostic specifications, the authors have tried to keep the method generalizable. It is possible that some, possibly newer OEMs, do not maintain even this level of detail. But even so, the new OEM stands to gain by following the conceptual approach behind this paper and the case study to ensure addressal of gaps in their concept.

The method proposed in this paper has only been applied in the case study from Section 5. The reason for choosing this particular subsystem as the target for this study, as discussed before, was because of the highly safety critical nature of the function of braking. Paradoxically, this could be seen a weakness in this paper as the level of rigour in the design of diagnostic specification might not exist in other traditionally non safety-critical components. Since automation can involve many different components, even traditionally non-safety critical components could become relevant to functional safety when automation is involved. For example, the ECU system responsible for controlling the headlights may have been assigned a lower ASIL level in a manually operated vehicle variant and may not have received equal care in monitor design. However, in the case of an automated driving scenario it could be a critical component if the cameras depend on the headlights for operational scenarios with low light. This method thus has a dependency on the quality of the diagnostic specification that is used as the input and may not be as effective for gathering information from certain components that have not been considered safety-critical before.

This paper does not claim to provide the one and only method to gather information needed for the design of the FSC; rather, it embraces the engineering design principle of sufficiency over completeness. It is used to design and justify the decisions made during the architecting of safety critical systems and gather evidence for safety argumentation. The method itself has only been tested in one scenario, and thus stands to benefit from another application in a longitudinal study to expand the claim of limited generalizability even further.

## 7. Related Work

To the best of our knowledge both the heuristics for removal of driver and the idea of using diagnostic specifications to aid in the process of automotive architectural design are novel. The proposed method however relates to several fields and research work and this section covers a sample of useful references for the reader for further detailed information.

Architectures for automated driving are a very popular topic at the time of writing of this paper. Several publications describe prototype autonomous vehicles including e.g. 'Bertha', the Mercedes Benz S-class vehicle [7], the European HAVE-IT project [8], Stanford University's DARPA Urban Challenge entry 'Junior' [9] and the A1 car, which won the 'Autonomous vehicle Competition', organized by Hyundai motors in 2010 [10], to name a few. Methodologies and reference architecture for vehicle control system design have also been proposed, e.g. by Gordon et al [11] and by Behere and Törngren in [12]. These methodologies include recommendations for layering for hierarchical control of various functionalities, and also compare and relate to well-known architecture control system patterns such as the subsumption architecture, NASREM [13] and 4D-RCS [14]. However these papers do not discuss the aspects of functional safety in detail and are more focused towards realizing the goal of automation. The mean time between vehicle crashes that cause injuries on the road in the US is about 50000 hours [15] and for an automated driving system to be acceptable to society, it is plausible that the system needs to be proven to be safer than this figure. Hence the challenge of functional safety in automated driving is more substantial than first apparent. A more detailed approach on the technical, practical and organizational challenges in architecting fully automated functionally safe driving can be found in [16].

The importance of early safety evolution in design is a well-known concept [17] and research has been done in the field of risk analysis and similar ideas as discussed in this paper, about functional failure propagation can be found in papers such as [18]. However in early phases of design, not many details are known about final implementation and requirements are constantly evolving requiring decisions to be made under uncertain information. Aerospace standards for functional safety such as ARP 4754A [19] take a continuous iteration between top-down design and platform constraints into account, place importance to the assumptions and decisions made with uncertain information during architectural design, and recommend tracking of these. While this rigour has not been part of ISO 26262, there has been some progress with methods such as those suggested in [20]

The effect of the lack of guidelines and the ambiguous definitions used for the PA is evident, with a short look into the academic and industrial papers from the automotive domain, in how the PA has been addressed. While we found many papers that evaluated the use of ISO 26262 on examples or on industrial case studies, no particular paper expanded upon how their PA was obtained. The PAs used in literature were found to be of varying levels of detail and content. Taylor et al. in [21] uses a hardware inspired PA, Westman and Nyberg in [22] use a pure software based PA, identify elements as software elements and leverage the information from their legacy example to make their case. [23] also includes the mechanical considerations such as installation space etc. [24] and [25] address the PA as functional blocks making no direct reference to a particular implementation technology. Though these papers do convey the ideas they were meant to convey, they do not explicitly account for legacy subsystems as does this paper.

The seminal work of Rechtin and Meier [26] describe the use of heuristics as a basic tool in architecting and works such as [27] have established the effectiveness of using heuristics and the prevalence of such methods. The authors agree with the view of Sexton [28] in that balance is needed in the speed of concept design and the assurance of functional safety and see the method proposed in this paper as a step forward to have a fast and inexpensive design concept without compromising on functional safety.

## 8. Conclusion

This paper presents a novel method to strengthen the process of developing the FSC by leveraging the information available in standard work products. The method allows for early, realistic decisions based on available facts and reduces the number of iterations in developing the FSC. The method allows for distributed development by allowing domain experts to highlight potential constraints and limitations in an effective manner for a class of functions using automation, allowing architects to view design trade-offs very early in the design phase thereby potentially reducing number of errors that can be made. The method allows for gauging what the platform is capable of and consequently allows for placing requirements on what the new components need to be developed. The intuitive heuristics (removing influence of driver) and the use of the standard work product (diagnostic specification) from the automotive industry, means that there is no specific training needed for the application of this method and that it is generalizable across OEMs and subsystems. This paper also demonstrates the application of the method in arguably one of the most safety critical subsystems in a vehicle i.e. the braking subsystem, to showcase its potential and discusses the promising results.

The novel contributions of this paper are as follows: (1) the idea of using of diagnostic specifications to facilitate the FSC design, (2) the analysis and the conceptual approach to FSC design, (3) the heuristics based method centred on the removal of the driver for functions related to automated driving and (4) the application of the concepts discussed to an industrial case study and the discussion.

To summarize, this paper thus takes a significant step towards requalification of legacy components and the design of the FSC using a strong grounding in already available information. With the rise of highly automated driving functions and the impending arrival of fully automated driving, the method presented in this paper allows for systematic architectural and safety design with cost savings.

### 8.1. Future Work

The use of diagnostic specifications and its explicit linkage to DTCs and failure reports from the field allow for future statistical analysis on the overall likelihood of failure of functions at a vehicular level. This statistical basis is targeted as future work. To refine the method itself, and to demonstrate its capabilities on other subsystems, a longitudinal case study on other safety critical subsystems at a passenger car OEM is also planned.

# Contact Information

Corresponding Author: Naveen Mohan; Email ID: naveenm@kth.se

# Acknowledgements

Support from FFI, Vehicle Strategic Research and Innovation and Vinnova through the ARCHER (proj. No. 2014-06260) project is acknowledged. The authors would also like to thank Per Roos and Johan Svahn from Scania for their help in reviewing this paper.